# About the domino problem in the hyperbolic plane from an algorithmic point of view

Maurice Margenstern,
Université Paul Verlaine − Metz,
LITA, EA 3097,
Île du Saulcy,
57045 METZ Cédex, FRANCE,
*e-mail*: *margens@univ-metz.fr*

May 24, 2018

**Abstract**

In this paper, we prove that the general problem of tiling the hyperbolic plane with *à la* Wang tiles is undecidable.

## 1 Introduction

The question, whether it is possible to tile the plane with copies of a fixed set of tiles was raised by Wang, [10] in the late 50's of the previous century. Wang solved the *partial* problem which consists in fixing an initial finite set of tiles: indeed, fixing one tile is enough to entail the undecidability of the problem. The general case, when no initial tile is fixed, was proved by Berger in 1964, [1]. Both Wang's and Berger's proofs deal with the problem in the Euclidean plane. In 1971, Robinson found an alternative proof of the undecidability of the general problem in the Euclidean plane, see [8]. In this 1971 paper, he raises the question of the general problem for the hyperbolic plane. Seven years later, in 1978, he proved that in the hyperbolic plane, the partial problem is undecidable, see [9]. Up to now, and as far as I know, the problem remained open.

In this paper, we give a proof that the general problem is also undecidable in the case of the hyperbolic plane. It is important to stress that in its spirit, our proof is a bit different from the proofs of Berger and Robinson. We consider the tiling process from the point of view of computer science. This means that it must be an algorithmic process which evolves in time.

We start from the initial idea of Berger's proof. It consists in simulating infinitely many computations of a Turing machine, indeed of the same Turing machine starting from an empty tape from which the set of prototiles is derived. The computations go as far as they can. If the Turing machine halts, the computations of the tiling arrive until the halting state is called and at this time, the tiling process is interrupted and it is no more possible to tile. If the Turing machine does not halt, the simulations go on endlessly or may be stopped by the limitations imposed on the tiling. But Berger, and after him Robinson, proved that among this infinity of computations in the case of a non-halting situation, either at least one of them can be performed without being interrupted, or there are arbitrary long computations. Accordingly the plane can be tiled. And so, tiling the plane is possible if and only if the Turing machine does not stop. And this proves the undecidability of the problem.

Analysing the proof, we can see that the restricted amount of space in the Euclidean plane forces to find a way to generate infinitely many bounded domains whose size is exponentially increasing. The exponential size is needed to overcome the meeting problem. It is also needed from Wang's remarks of the 50's: if tilings of the plane are necessarily periodic, the general tiling problem is decidable. And, historically, Berger's proof had, as a side-product, the construction of a non-periodic tiling. This was not the goal of Berger's paper, but this was entailed by Wang's remark. Berger managed to produce exponential signals in a rather easy way. Also, the exponential growth guarantees the existence of at least an infinite sequence of increasingly longer and longer computations. It is important to focus on the fact that the constructions performed both by Berger and by Robinson in their respective proofs make a huge use of similarity. In his 1978 paper, speaking of the problem in the hyperbolic plane, Robinson remarks that *We cannot imitate Penrose construction, since similarity is impossible in the hyperbolic plane.* This remark extends to Berger's proof and also to 1971 Robinson's proof.

The situation in the hyperbolic plane is very different from the Euclidean case, quite the opposite: see [5] for a more detailed description. Already Robinson's proof of the partial case witnesses at the key point: in the hyperbolic plane, orientation and localisation are very difficult. This is what is expressed by Robinson in his 1978 paper: *The group of motions in the hyperbolic plane does not have a uniquely defined subgroup which plays the* *same role as the group of translations in the Euclidean plane.* Indeed, the group of direct motions of the hyperbolic plane is simple which explains the impossibility of finding a *nice* subgroup as indicated by Robinson. Still in this 1978 paper, Robinson makes an interesting remark, asking whether *the undecidability and non-periodicity results about tilings of the Euclidean plane have analogs for the hyperbolic plane*: *It is no longer clear that the two problems are related.* And it can be noticed that the non-periodicity problem for tilings of the hyperbolic plane has drawn more attention than the undecidability problem, see for instance [6, 2] for works in this direction.

From the tools I devised to locate cells of a cellular automaton implemented in the hyperbolic plane, see [4, 5] for more complete references, it came to me that perhaps, this could provide us with a new angle to tackle the problem.

Due to the very different context, an important point of the discussion is the exact meaning of the general problem. It cannot be the exclusion of any constraints on the initial tiling set and on the construction. Indeed, if the first tile can be chosen at random, it can bear the computation signs of a Turing machine. In this case, it may turn out that this tile will never be used, and it should not be placed. But the reachability problem is undecidable for Turing machines. Consequently, the organisation of the tiling must force any computation to be started at the very beginning of the Turing computation. And so, the first tile cannot be taken at random. In his proof, Berger defines a subset of the initial set of tiles, the **skeleton**, from which the first tile is chosen. Here also, we define such a skeleton and, as in Berger's proof, it allows to construct the general pattern in which areas are delimited for the Turing computation.

Here, we suggest to give a precise definition of the partial case and of the general case. The partial problem means that at least one tile is fixed in the initial set which must be used but at most finitely many times. In the general problem, there is a subset of the initial set, the **skeleton**, whose elements must all be used infinitely many times in the tiling.

Our solution is **algorithmic**. The undecidability of the tiling problem has no meaning for solutions which would not be algorithmic. As in Berger's and Robinson's proof, our set of initial tiles has the property that there is an algorithm which produces any possible solution by running the algorithm non-deterministically. In Berger's and Robinson's proof, the initial tiles can be only translated when they are duplicated to be placed in the tiling. Rotations and reflections are ruled out. Here, we rule out only reflections: from the negative curvature of the hyperbolic plane, we know that the translations which leave the tiling invariant also generate the rotations which leave the tiling invariant.

Our solution sticks to Wang's convention on the tiles as in Berger's paper. Also, in the case when the Turing machine does not halt, we have infinitely many **periodic** solutions. This confirms the quoted above prediction of Robinson. We also prove that are there countably many periodic solutions which, indeed, can be enumerated.

In this abstract, we have no room to remind the reader of what is needed from hyperbolic geometry. It is also impossible to redefine the tools I introduced for the location of tiles in the hyperbolic plane. The reader is invited to look at the technical report, [5] on which the paper is based and which is available at the following address:

`http://www.lita.sciences.univ-metz.fr/~margens/hyp_dominoes.ps.gzip`

In the second section, the paper constructs the tiling defined by the skeleton, the **mantilla**. In the third section, it presents the partial solution, within the frame of the **harp**. In the fourth section, it presents the general solution.

# 2 The mantilla

This section addresses a particular algorithmic construction which takes place in the tiling of the hyperbolic plane which we call the **ternary heptagrid**, see [5] also for further references, and which is usually known as $\{7,3\}$, see figures 5 and 6, further.

It is generated by the regular heptagon with vertex angle $\frac{2\pi}{3}$ by reflections in its edges and, recursively, of the images in their edges.

## 2.1 The flowers

In the ternary heptagrid, a **ball** of **radius** $n$ around a tile $T_0$ is the set of tiles which are within distance $n$ from $T_0$ which we call the **centre** of the ball. The **distance** of a tile $T_0$ to another $T_1$ is the number of tiles constituting the shortest path between $T_0$ and $T_1$. As we shall be very often concerned by balls of radius 1 only, we give them a special name, we call them **flowers**.

The mantilla consists in merging flowers in a particular way. It comes from the following consideration. Robinson's proof of the undecidability of tiling the Euclidean plane is based on a simple tiling consisting of two tiles represented by the left-hand side of figure 1, below.

On the right-hand side of the figure, we have the 'literal' translation of these tiles in the ternary heptagrid. It is not difficult to see that putting three copies of $a$ around $b$ leads to an impossibility. However, a slight modification of the tile $b$, see the tile $c$ in figure 2, leads to the solution.

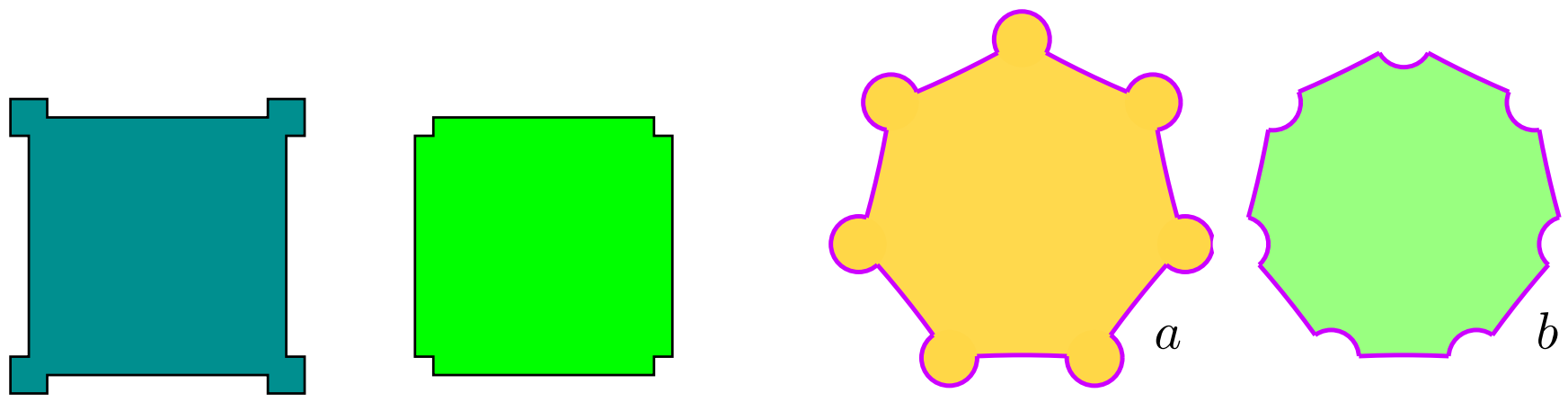

**Figure 1** *On the left: Robinson's basic tiles for the undecidability of the tiling problem in the Euclidean case. On the right: the tiles a and b are a 'literal' translation of Robinson's basic tiles to the situation of the ternary heptagrid.*

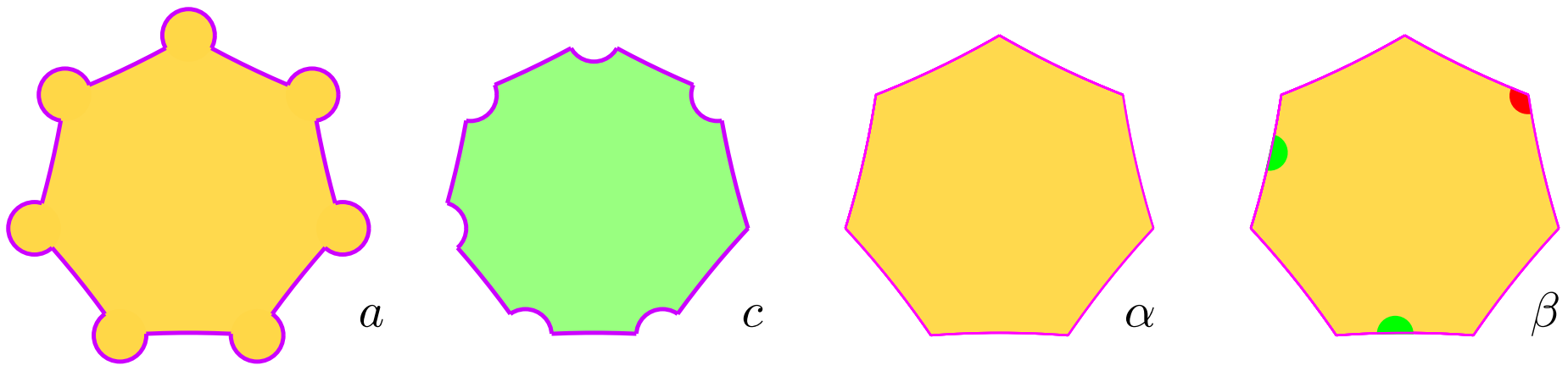

**Figure 2** *On the left: change in the tiles à la Robinson. On the right: their translation in pure Wang tiles.*

On the right hand side of figure 2, we have the transformation of tiles $a$ and $c$ into Wang tiles. We call the tile $\alpha$ a **centre** and the tile $\beta$ a **petal**. As represented in figure 2, centres and petals are not enough, as centres alone may tile the hyperbolic plane, we shall fix this in section 2.2. From now on, a **flower** is the figure consisting of a centre surrounded by petals.

It is not difficult to see that there can be several types of flowers, considering the number of red vertices for which the other end of an edge is a vertex of a centre. We say that the red vertex is at distance 1 of this centre. Now, a flower may have 3, 2, 1 or 0 red vertices at distance 1. This corresponds to the fact that, respectively, 7, 8, 9 or 10 centres can be put around the considered flower. We shall respectively speak of **i**-flowers for $i \in \{7..10\}$. Now, there are two cases of **9**-flowers, depending on the smallest number of petals between the two red vertices: 2 or 3. They will be called $F$- and $G$-flowers, respectively. We shall prove that using $F$-, $G$- and **8**-flowers only, we can tile the hyperbolic plane using the tiles $\alpha$ and $\beta$.

## 2.2 The tiles for the mantilla

The next ingredient consist in putting **numbers** on the edges of the centres which can be seen as different colours of the edges. This is needed to fix an algorithmic construction of the tiling. Also, the numbering prevents to tile the plane with centres only. Figure 3 gives a few samples of the tiles.

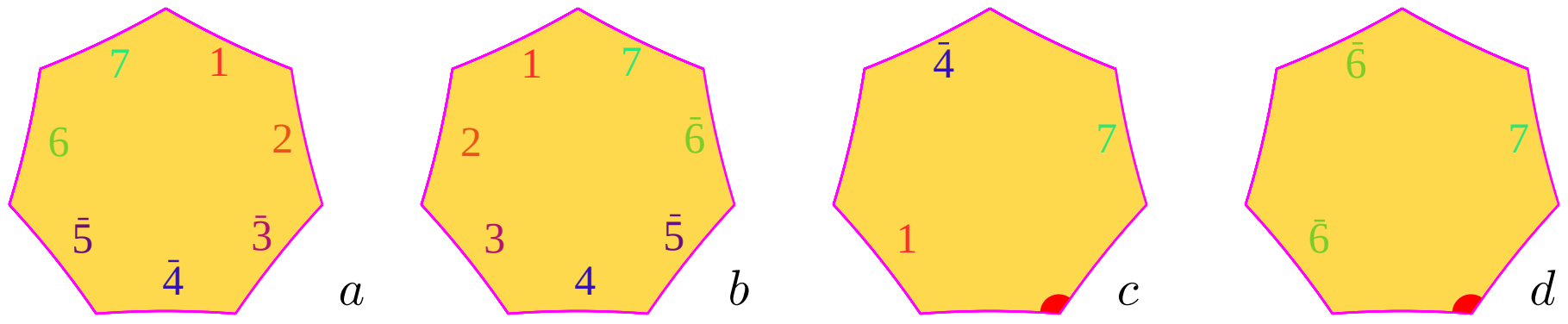

**Figure 3** *Samples of tiles for the mantilla: an $F$-centre, a $G_\ell$-centre and two petals:* $1\overline{4}7\circ$ *and* $\overline{6}\overline{6}7\circ$*, see table* 2*, below.*

The explanation of the numbering is given both by figures 3 and 4. We start with an $F$-flower. From the red vertices, we draw rays: the ray supporting the edge shared by petals 13-14 on the left-hand side; the ray supporting the edged shared by petals 16-17 on the right-hand side. We define a similar sector for $G$-flowers, again based on the red-vertices. For an **8**-flower, we take the union of four $F$-sectors induced by the petals below the **8**-centre. As shown by figure 4, the sectors can be split into sub-sectors of the three types: $F$-sector, $G$-sector or half of an **8**-sector, following the splitting technique of [4].

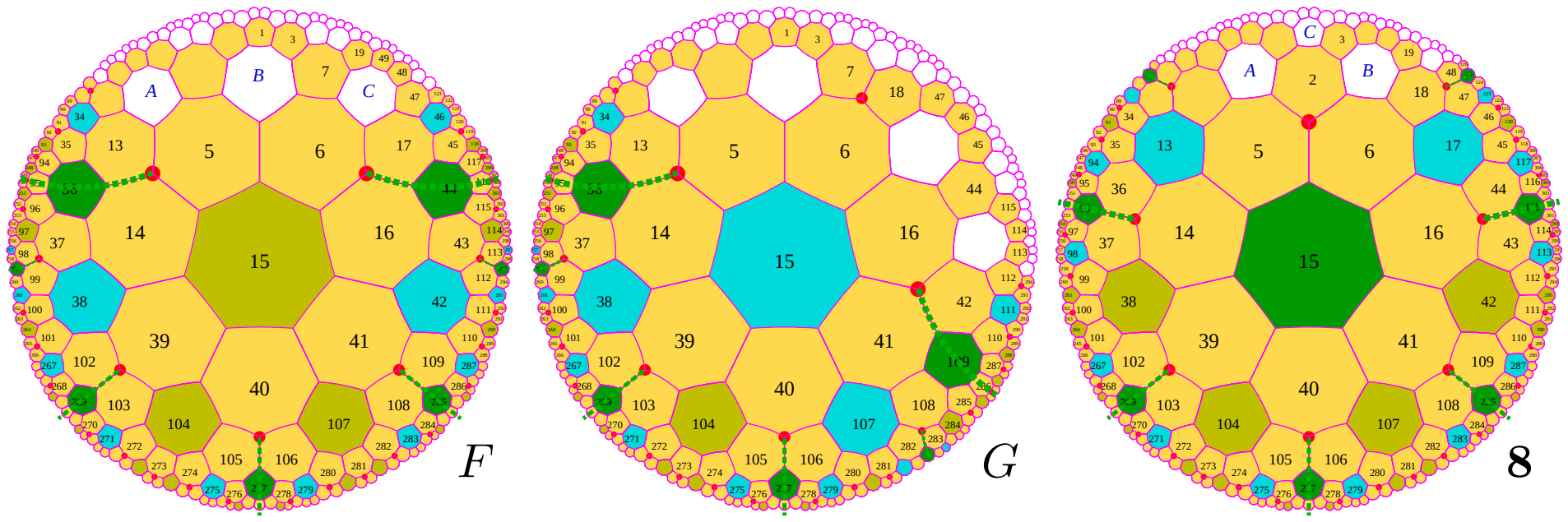

**Figure 4** *Splitting of the sectors defined by the flowers. From left to right: an* $F$*-sector,* $G$*-sector and* **8***-sector.*

Following the technique recalled in [4], the splitting depicted by figure 4 defines a tree. Accordingly, we shall later speak of $F$- and $G$-sons of flowers or centres, when it will be convenient to identify them.

The numbering is first defined for centres. It goes clockwise around the centre for $F$- and **8**-centres and counter-clockwise for $G$-centres. The numbering also allows to differentiate $G$-flowers into $G_\ell$- and $G_r$-flowers: they appear on the left-hand or right-hand side in the splitting. In order to better differentiate centres, we also overline numbers. The numbering of centres induce the numbering of the petals according to the following tables. In the table, with respect to a centre, a petal is called **non-parental** if it abuts at an edge $\mathbf{i}$ of the centre with $\mathbf{i} \notin \{1, 7\}$. Otherwise, it is **parental**.

| | 2 | 3 | 4 | 5 | 6 |
|---|---|---|---|---|---|
| $F$ | 2 | $\overline{3}$ | $\overline{4}$ | $\overline{5}$ | 6 |
| $G_\ell$ | $\overline{6}$ | $\overline{5}$ | 4 | 3 | 2 |
| $G_r$ | 6 | 5 | 4 | $\overline{3}$ | $\overline{2}$ |
| **8** | $\overline{2}$ | 3 | $\overline{4}$ | 5 | $\overline{6}$ |

**Table 1** *Table of the distribution of colours on the sides of the central tiles. Labels* 1 *and* 7 *are not indicated: they are the same for $F$- and* **8***-flowers, and they are exchanged for $G$-flowers.*

It can be noticed that table 2 fixes the non-parental petals of a centre. Also, the parental petals can be chosen freely within fixed sets of possibilities, see [5] for a precise description.

Accordingly:

**Lemma 1** *The* 21 *tiles of tables* 1 *and* 2 *allow to tile the hyperbolic plane. Moreover, there is an algorithm to perform the tiling which we call the* **mantilla***.*

| | 2 | $\overline{2}$ | 3 | $\overline{3}$ | 4 | $\overline{4}$ | 5 | $\overline{5}$ | 6 | $\overline{6}$ |
|---|---|---|---|---|---|---|---|---|---|---|
| $F$ | $2\circ77$ | | | $1\circ1\overline{3}$ | | $1\overline{4}7\circ$ | | $\overline{5}7\circ7$ | $11\circ6$ | |
| $G_\ell$ | $11\circ2$ | | $37\circ7$ | | $1\circ14$ | | | $\overline{5}\circ77$ | | $\overline{66}7\circ$ |
| $G_r$ | | $1\overline{22}\circ$ | | $11\circ\overline{3}$ | $47\circ7$ | | $1\circ15$ | | $6\circ77$ | |
| **8** | | $1\overline{22}\circ$ | $137\circ$ | | | $1\overline{4}7\circ$ | $157\circ$ | | | $\overline{66}7\circ$ |

**Table 2** *Table of the non-parental petals according to their parent flower. The numbering of a tile is established by starting from the smallest number and running clockwise around the petal. Symbol $\circ$ indicates where is the red vertex between the numbers.*

# 3 The partial problem: the harp

Our general solution relies on a solution of the partial problem which is different from Robinson's construction of [9]. In this paper, two tilings are used, a hexagonal one and a quadrangular one, the dual graph of the pentagrid $\{5, 4\}$.

Here, we define the solution in an **angular sector**, see [5]. Such an angular sector is represented on the left-hand side of figure 5 by the two thick yellow rays supported by mid-point lines, *i.e.* lines passing through mid-points of pairwise consecutive edges of a heptagon.

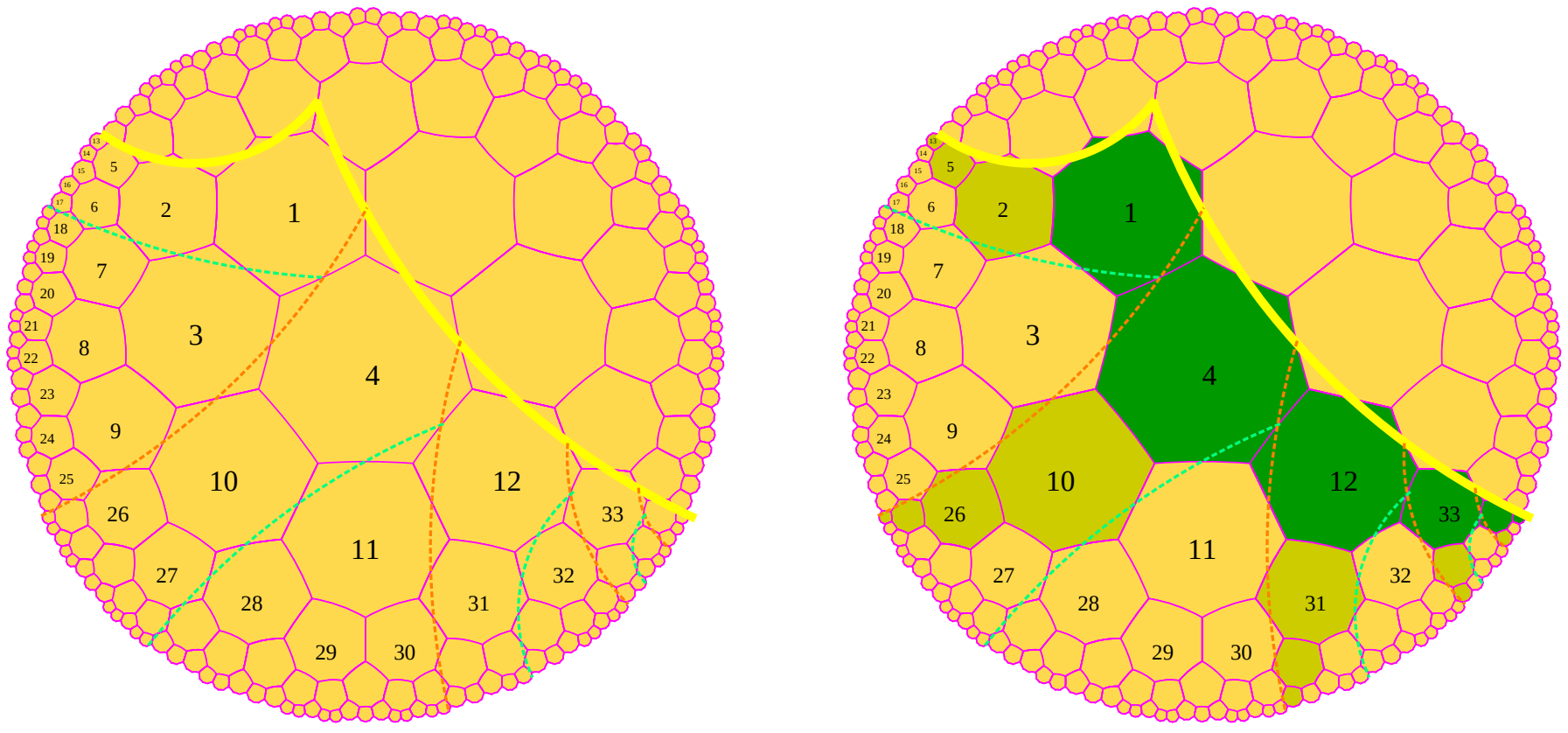

**Figure 5** *The guidelines for the harp.*

As indicated in [5], a standard Fibonacci tree, see [4, 5] for this notion and the corresponding technique, can exactly be inserted in such an angular sector.

On the right-hand side of the figure, we can see the **harp** itself. It contains a Fibonacci tree which can be viewed as a space time diagram of the Turing computation. The rightmost branch of the tree, the dark green tiles on the right-hand side of figure 5, call it the **frame**, represents the Turing tape at the initial time. We may assume that the tape is empty at initial time and, without loss of generality, that the head never goes to the left of a fixed cell, here the root of the tree. The evolution in time of one cell is the leftmost branch of the Fibonacci subtree rooted at the image of this cell on the frame. The head moves from cell to cell running on **levels** of the Fibonacci tree, see [5], between two cells. A few tiles are enough for this purpose, see [5] for an exact description.

# 4 The general problem

## 4.1 Refining the mantilla

The idea is to introduce trees in which harps will be inserted. For this purpose, for each $G$-flower, we decide to construct a tree with the centre of its $F$-son as a root. Call such a tree a **candidate** and the set of tiles which it contains its **area**.

The key property, which is proved in [5] is that the areas of two candidates are either disjoint or one of them contains the other. This allows to define the notion of **thread** of candidates, a linearly ordered set of candidates containing all the candidates between two of them. There may be also an **ultra-thread**, a maximal thread with respect to inclusion. Some mantillas do possess ultra-threads and others do not. We consider only mantillas without ultra-threads. It is not difficult to prove that there are continuously many of them. In such a mantilla, each thread has a maximal element which we call a **selected tree**. Next, we remove the areas of all selected trees and what remains is called the **shrunken** mantilla.

As done in [5], it is not difficult to prove that the shrunken mantilla is an infinite connected set of tiles. Also, the ball of radius 6 around any tile of the shrunken mantilla contains a $G$-centre. This shows that there are infinitely many roots of selected trees and that they constitute a rather dense set of the shrunken mantilla.

The shrunken mantilla no more tiles the hyperbolic plane. But its complement in the plane is exactly the union of all selected trees which pairwise have disjoint areas. Now, we are free to restore the trees as we wish. In each of them, we insert a harp as indicated in the right-hand side of figure 6. The new tiling is called the **refined mantilla**. The shrunken mantilla is really a skeleton, delimiting the computation regions **outside** itself. In the Euclidean case, Berger's and Robinson's constructions define the computation regions **inside** the tiling obtained from the skeleton.

The pictures of figure 6 give an idea of the transition tiles between the different areas which are needed for the final set of tiles. In [5], we very accurately describe a set of 64 tiles to construct any refined mantilla, 29 ones of them constituting the skeleton. Let us indicate that this skeleton consists of the 21 tiles of section 2.2 except for tiles $37\circ7$, $1\circ14$, $47\circ7$ and $1\circ15$ which now bear the marks of the selected tree, see the left-hand picture of figure 6 and a half of the tiles of the border with the selected tree, see [5]. Also, 19 tiles bear the computation signs. Now, to each Turing machine working in the conditions fixed in the section devoted to the harp, we define a set of tiles where, by duplicating the 19 computing prototiles, we obtain all possible combinations with the actual signs of the considered Turing machine. Then we have an algorithm to construct the tiling.

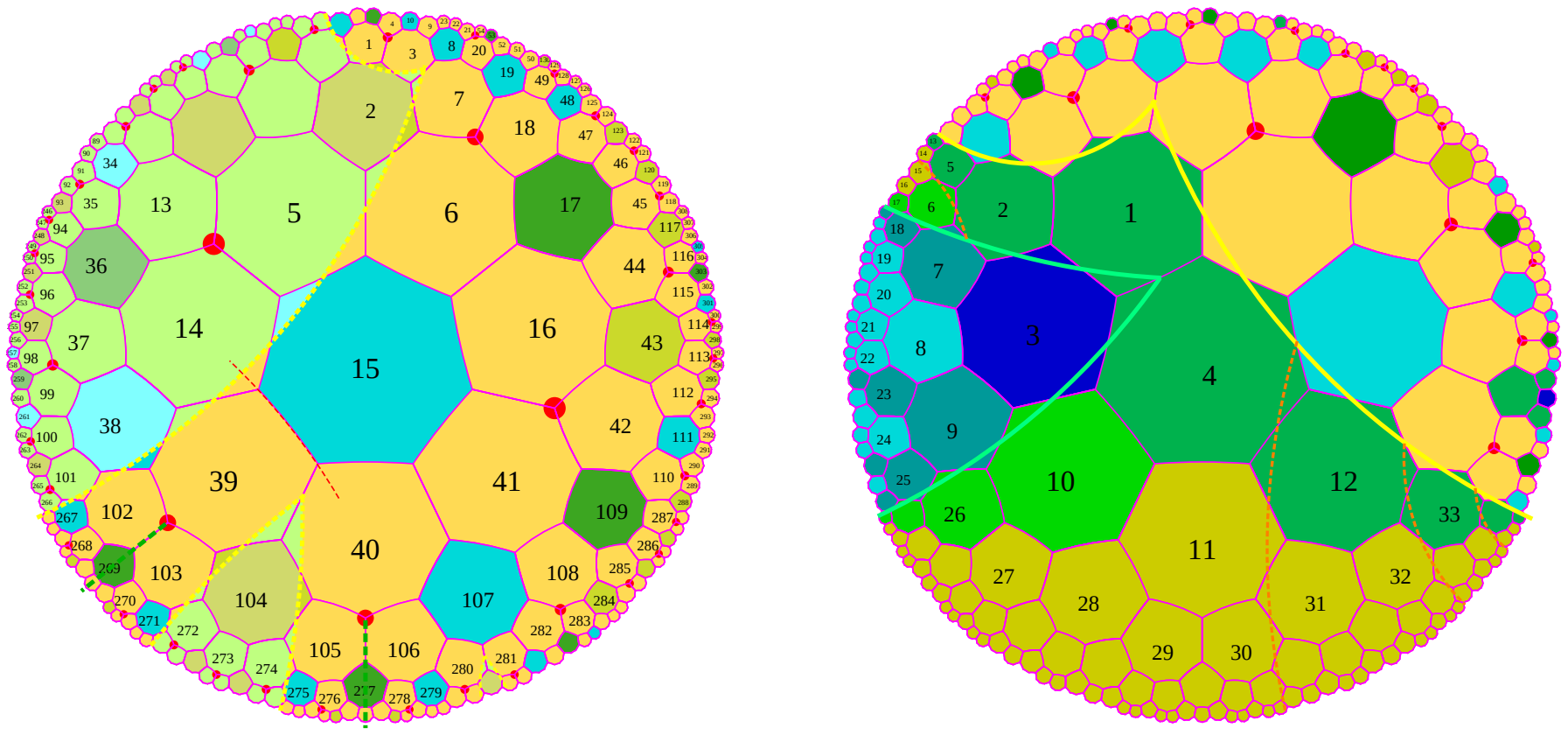

**Figure 6** *On the left, two selected trees and the shrunken mantilla around them. On the right, the insertion of the harp, in blue, inside the selected tree, in green. On the right-hand side of the selected tree with numbered tiles, notice another selected tree at three tiles on the right-hand side of tile* 33*. This new selected tree is generated by the* $G_\ell$*-centre which is adjacent to tiles* 4 *and* 12*.*

First, a tile of the skeleton is taken at random. This amounts to choose a tile of the shrunken mantilla at random. After this, tiles are chosen in the full set and once a tile is placed, it can never be removed. The working of the algorithm is split into times $t_n$, $t_0$ being the choice of the first tile $T_0$. Between time $t_n$ and $t_{n+1}$, the algorithm extends the previous area $\Sigma_n$ reached at time $t_n$. $\Sigma_n$ is headed by a flower without its parental tiles and it is assumed to contain a ball of radius $n$ around $T_0$. The algorithm chooses the parental tiles at random among what is possible. Then, it goes on the construction of a new sector, again headed by a flower without parental tiles until it contains strictly $\Sigma_n$ and a ball or radius $n+1$ around $T_0$. When this is performed, $\Sigma_{n+1}$ and time $t_{n+1}$ are reached.

In the random choice, it may happen that the algorithm takes a tile which is the border between the shrunken mantilla and a selected tree. Here, we can see why the insertion of the harp as displayed by figure 6 is performed in a subtree inside the selected tree: it allows to avoid a computing tile when the algorithm performs a choice at random.

This proves the following theorem:

**Theorem 1** *There is a set of* 64 *tiles which allows to construct any refined mantilla up to standard transformations for a Turing computation. The* *above algorithm allows to obtain any solution when the Turing machine does not halt, and there are continuously many of them. Among these solutions, countably many of them are periodic.*

**Corollary 1** *The general tiling problem is undecidable for the hyperbolic plane.*

Adapting the construction of [3, 7] to the harp as indicated in [5] we get:

**Corollary 2** *There is a finite set $\mathcal{S}$ of tiles such that there is a non-recursive way to tile the hyperbolic plane with copies of $\mathcal{S}$ but no recursive way.*